
\vsize 8.7in
\def\doublespace{\baselineskip 22.76 pt}
\hsize 6.5 true in
\hoffset=0. true in
\voffset= 0. true in
\def\mathnew{\mathsurround=0pt}
\def\simov#1#2{\lower .5pt\vbox{\baselineskip0pt \lineskip-.5pt
\ialign{$\mathnew#1\hfil##\hfil$\crcr#2\crcr\sim\crcr}}}
\def\simgreat{\mathrel{\mathpalette\simov >}}
\def\simless{\mathrel{\mathpalette\simov <}}
\font\twelverm=cmr10 scaled 1200
\font\ninerm=cmr7 scaled 1200
\font\sevenrm=cmr5 scaled 1200
\font\twelvei=cmmi10 scaled 1200
\font\ninei=cmmi7 scaled 1200
\font\seveni=cmmi5 scaled 1200
\font\twelvesy=cmsy10 scaled 1200
\font\ninesy=cmsy7 scaled 1200
\font\sevensy=cmsy5 scaled 1200
\font\twelveex=cmex10 scaled 1200
\font\twelvebf=cmbx10 scaled 1200
\font\ninebf=cmbx7 scaled 1200
\font\sevenbf=cmbx5 scaled 1200
\font\twelveit=cmti10 scaled 1200
\font\twelvesl=cmsl10 scaled 1200
\font\twelvett=cmtt10 scaled 1200
\skewchar\twelvei='177 \skewchar\ninei='177 \skewchar\seveni='177
\skewchar\twelvesy='60 \skewchar\ninesy='60 \skewchar\sevensy='60
\def\twelvepoint{\def\rm{\fam0 \twelverm}
  \textfont0=\twelverm \scriptfont0=\ninerm \scriptscriptfont0=\sevenrm
  \rm
  \textfont1=\twelvei \scriptfont1=\ninei \scriptscriptfont1=\seveni
  \def\mit{\fam1 } \def\oldstyle{\fam1 \twelvei}
  \textfont2=\twelvesy \scriptfont2=\ninesy \scriptscriptfont2=\sevensy
  \def\cal{\fam2 }
  \textfont3=\twelveex \scriptfont3=\twelveex \scriptscriptfont3=\twelveex
  \textfont\itfam=\twelveit \def\it{\fam\itfam\twelveit}
  \textfont\slfam=\twelvesl \def\sl{\fam\slfam\twelvesl}
  \textfont\bffam=\twelvebf \scriptfont\bffam=\ninebf
    \scriptscriptfont\bffam=\sevenbf \def\bf{\fam\bffam\twelvebf}
  \textfont\ttfam=\twelvett \def\tt{\fam\ttfam\twelvett}
  }
\def\folio{\ifnum\pageno=1\nopagenumbers\else\number\pageno\fi}

\font\twelvess=cmss10 scaled 1200
\twelvepoint
\doublespace
\def\ref{\par\noindent\hangindent=2pc \hangafter=1 }
\def\apj{{\it Ap.~J.}}
\def\apjl{{\it Ap.~J. (Letters)}}
\def\cm{\hbox{cm}}
\def\s{\hbox{s}}
\def\ergs{\hbox{ergs}}
\def\G{\hbox{G}}

\def\MeV{\hbox{MeV}}
\def\b{\hbox{\bf B}}
\def\e{\hbox{\bf E}}

\null
\centerline{Submitted to the Editor of the Astrophysical Journal
{\sl Letters} on May 12, 1993}
\vskip 0.1in
\centerline{Revised June 10, 1993}
\null\vskip 0.85 true in
\centerline{\bf FOCUSING OF ALFV\'ENIC WAVE POWER IN THE}
\vskip 0.1in
\centerline{\bf CONTEXT OF GAMMA-RAY BURST EMISSIVITY}
\vskip 1.0 true in
\centerline{{\bf Marco Fatuzzo}\footnote
{\hbox{$\null^{\dag}$}}{Compton GRO Fellow.}}
\vskip 0.05in
\centerline{\sl Department of Physics, University of Michigan}
\centerline{\sl Ann Arbor, MI 48109}
\vskip 0.05in
\centerline{and}
\vskip 0.05in
\centerline{{\bf Fulvio Melia}\footnote
{\hbox{$\null^*$}}{Presidential Young Investigator.}}
\vskip 0.05in
\centerline{\sl Department of Physics and Steward Observatory,
University of Arizona}
\centerline{\sl Tucson, AZ 85721}
\vfill\eject
\null
\centerline{\bf Abstract}
\bigskip
{\twelvess

Highly dynamic magnetospheric perturbations in neutron star
environments can naturally account for the features observed
in Gamma-ray Burst spectra.  The source distribution, however,
appears to be extragalactic.  Although noncatastrophic
isotropic emission mechanisms may be ruled
out on energetic and timing arguments, MHD processes can produce
strongly anisotropic $\gamma$-rays with an observable flux out
to distances of $\sim 1-2$ Gpc.  Here we show that
sheared Alfv\'en waves propagating along open magnetospheric
field lines at the poles of magnetized neutron stars transfer their
energy dissipationally to the current sustaining the field
misalignment and thereby focus their power into a spatial region
$\sim 1000$ times smaller than that of the crustal disturbance.
This produces a strong (observable) flux enhancement along certain directions.
We apply this model to a source population
of ``turned-off'' pulsars that have nonetheless retained their
strong magnetic fields and have achieved alignment at a period of
$\simgreat 5$ seconds.

}
\bigskip\noindent {\it Subject headings}:  acceleration of particles --
cosmology: miscellaneous -- galaxies: evolution -- gamma rays: bursts --
MHD -- pulsars: general

\vfil\eject
\centerline{\bf 1. Introduction}
\medskip

The lack of a precise determination of a distance scale to Gamma-ray
burst (GRB) sources has greatly hindered our theoretical understanding
of these objects.  Much of what we know about these bursts is based
on inferences drawn from clues provided by their spectra, including:  (1)
Most events exhibit rapid variability, apparently on time scales
shorter than the best available instrument temporal resolution,
with one burst exhibiting structure on
a $200\;\mu$s timescale (Bhat et al. 1992).  This variability indicates that
the sources must be very compact, characterized by a length scale no
larger than $\sim 50-100$ km.  (2) Bursts can last anywhere from tens of
milliseconds to as long as 900 seconds, and usually have a complex temporal
structure. This complicated time dependence would seem to
favor mechanisms that invoke highly dynamic perturbations in otherwise
stable environments. (3) Typical GRBs emit a substantial fraction  of their
power at energies in excess of $\sim 1$ MeV, which suggests that
nonthermal processes are responsible for the emission of the
$\gamma$-rays. In this regard, the inverse-Compton scattering of soft
photons by relativistic particles has been very successful in reproducing
the observed spectra (e.g., Pozdnyakov, Sobol' \& Sunyaev 1977; Canfield,
Howard \& Liang 1987; Melia \& Fatuzzo 1989; Ho \& Epstein 1989; Melia
1990a,b).

It is possible to account for these observations by invoking a model in
which the bursts originate within the magnetosphere of strongly
magnetized neutron stars. In addition to being very compact, these
environments are subject to magnetic fluctuations on sub-millisecond
timescales, and the constituent particles can be energized nonthermally
via the induced electrostatic forces.

As is well known, however, the neutron-star paradigm must be reconciled
with the uniform, yet spatially truncated GRB distribution observed by the
BATSE experiment on CGRO (Meegan et al. 1992).  These observations
seem to rule out nearby (i.e. Galactic) single population models, and have
therefore led to renewed speculation that GRBs originate at cosmological
redshifts.  But a naive estimate of the burst energy
required for such distant sources yields a value that is significantly
larger than that which a neutron star could reasonably produce, unless the
event was catastrophic (e.g., the coalescence of a neutron-star binary,
Narayan et al. 1992), which does not seem to be borne out by the time
history of typical bursts.

A resolution to this apparent conflict was recently proposed by Melia \&
Fatuzzo (1992, hereafter MF), in which sheared Alfv\'en waves generated
near the polar cap of strongly-magnetized neutron stars produce streams
of relativistic particles that are focused by the underlying magnetospheric
structure.  These energetic charges upscatter the radio-frequency photons
(emitted at larger radii) into $\gamma^{-1}$ cones aligned with the
underlying magnetic field lines, resulting in an enhanced $\gamma$-ray
flux along preferred lines of sight.  This anisotropic emission is such that
a pulsar glitch releasing $\sim 10^{45}$ ergs of energy could be viewed as
a GRB out to a distance of $\simgreat 1$ Gpc.  A key assumption of this
scenario is that the Alfv\'enic power can indeed be emitted
anisotropically.  We show in this Letter that the required focusing is a
natural consequence of the dissipational properties of sheared Alfv\'en
waves whose shear lengthscales ($s\simless 10$ cm) are much
smaller than the size of the region ($\simgreat 10^4$ cm) encompassing
the overall Alfv\'en wave fluctuation.  As such, this work strengthens the
case for a non-catastrophic, cosmological origin of GRBs, and supports our
view that an improved understanding of the micro-physical processes in
neutron-star environments can indirectly, though significantly, influence
our study of galactic evolution out to a redshift in excess of $1-2$
(Tamblyn \& Melia 1993).

\medskip
\centerline{\bf 2. Sheared Alfv\'enic Wave Dissipation}
\medskip

The general theory of sheared Alfv\'en waves (SAWs) has been developed
elsewhere (Melia \& Fatuzzo 1992; Fatuzzo \& Melia 1993).
Here we consider the global properties of their dissipation.  We idealize
the unperturbed polar cap region as a fully ionized, homogeneous plasma
threaded by a uniform magnetic field $\b_0 = B_0\hat z$.  The sheared
Alfv\'en waves may therefore be described by magnetic perturbations of the form
$$
\b_A = B_a(y)\;\exp(ikz-i\omega t)\hat x\;,\eqno(1)
$$
where $B_a(y)$ is an odd function that characterizes the shear geometry.
For our purposes here, we take the shear profile to be
$$
B_a(y) = \cases{  B_{a0}&$S\ge y \ge s$\cr
\null&\null\cr
B_{a0}\; g(y)
&$ |y|\le s$\qquad\quad,\cr
\null&\null\cr
-B_{a0}&$-S\le y\le -s$}\eqno(2)
$$
where $g(y)$ is a continuous function that satisfies
the condition $g(\pm s)=\pm1$, and where $s\ll S$
($s$ being the shear lengthscale and $S$ the lengthscale
of the encompassing plane wave regions).
For convenience, we define $\eta$ as the ratio of the
nonsheared to sheared surface areas ($\eta\equiv S/s$ in the present geometry).

It is clear from the form of $B_a(y)$ and Amp\'ere's Law that an
electric field $E_{Az}$ must exist inside the sheared region
($|y|<s$) until a sufficiently strong current $J_s$ is produced parallel to
the underlying magnetic field $\b_0$. For these waves, the equilibrium
Goldreich-Julian particle density $n_0$ is insufficient to support a
current large enough to short out $E_{Az}$. Charges must therefore be
copiously stripped off the stellar surface, thereby inducing a charged
particle flow to give the required $J_s$.  Since the Alfv\'en speed
is $v_\alpha\gg c$, the waves travel with a phase velocity
$u_\alpha = \omega/k\approx c$.  In order for $J_s$ and the
encompassing magnetic shear to remain in phase, the particle flow
must be relativistic, and thus, have an average density
$$
n_s\approx {B_{a0}\over 4\pi e s}\approx 1.7\times 10^{19}\cm^{-3}\;
\left({B_{a0}\over 10^{12}\;\G}\right)\;
\left({s\over 10\;\cm}\right)^{-1}\;.\eqno(3)
$$
We note that the stripped particles escape from the system by flowing
out along the open magnetospheric field lines.

In standard pulsar theory, radio emission results from the coherent
motion of ``bunches'' of electrons streaming along open field lines
with Lorentz factors $\gamma\sim 10^{4-5}$. As such, strong transient
radio emission is expected to be a natural byproduct of sheared
Alfv\'en waves if similar particle energies are reached, and if
this emission is produced with front-back symmetry along
the local field-line direction, a large fraction of the overall
radio flux will naturally be funneled back onto the polar cap.
Taking into account the coherent nature of the processes responsible
for pulsar emission, we parametrize the flux impinging onto the
stellar surface by $F_r = \xi\eta^{-2} (n_s/n_C)^2 \;(L_C/\pi R_{pc}^2)$,
where $L_C$ and $n_C$ are the Crab pulsar luminosity and magnetospheric
number density, respectively, and where $R_{pc}$ is the radius
of the open field line polar cap.  Assuming a stellar radius of
$R_*=10^6\cm$, $R_{pc}$ can be related to the pulsar period
$P$ via $R_{pc} = 1.4\times 10^4\;\cm\;(P/1 \s)^{-1/2}$.
With $L_C=10^{32}\;\ergs\;\s^{-1}$ and $n_C =10^{13}\;\cm^{-3}$, this yields
$$
F_r= 2\times 10^{30} \ergs\;\cm^{-2}\;\s^{-1}\;\xi
\left({\eta\over 10^3}\right)^{-2}\;
\left({B_{a0}\over 10^{12}\;\G}\right)^2\;
\left({s\over 10\; \cm}\right)^{-2}\;
\left({P\over 5\;\s}\right)\;,\eqno(4)
$$
where $\eta^{-1}$ is the sheared flow ``filling factor''.
The parameter $\xi$ encompasses both geometric and emission
uncertainties, and as such is poorly known.  We note that
if $\xi$ becomes too small ($\simless 0.1$ for the range of
parameters considered here), the wave dissipation
lengthscale due to field line annhilation becomes much larger
than $R_*$, and the SAW mechanism becomes inefficient at
producing $\gamma$-rays (see the discussion after
equation [7]).  However, since the Crab pulsar is itself very
inefficient at converting spin-down energy into radio
emission compared to typical pulsars, and since
we have made the conservative assumption that the (coherent)
radio flux scales as $\eta^{-2}$ (i.e., the square of the {\it total} number
of particles), it is reasonable to assume that $\xi\gg 0.1$ (see also
the discussion after equation [11]).

The presence of $F_r$ results in a radiative drag on the relativistic
current-carrying charges.  By analogy with MHD phenomena, the current driving
electric field ($\e=E_{Az}\hat z$) must be generated within the
shear at the expense of the magnetic wave energy. However, the
sheared waves are distinguished from pure MHD
fluctuations for two important reasons.  First, the charges
which generate $J_s$ are constrained
to always move along the same $\b_0$ field lines, so that SAWs
cannot easily change their initial structure. Second, the simple
concept of Ohm's law is not valid for the relativistic flow inside
the shear.  Indeed, once the particles become relativistic,
the current quickly
decouples from the driving electric field,  and since the radiative
drag increases rapidly with $\gamma$ (the particle Lorentz factor),
one might expect that a mildly relativistic flow will be favored
by the system.

Though $E_{Az}$ depends on the microphysics of the shear (including all the
annihilation processes, such as the tearing mode instability), its
value may be estimated with a relatively simple argument under the
assumption that the annihilation time scale within the shear is the
shortest of the relevant time scales.  The strength of the electric
field is limited by the rate at which the oppositely-directed magnetic
fluctuations are driven together by the large magnetic pressure
gradients associated with SAWs.  Since the Alfv\'enic field lines are
strongly coupled to $B_0$ via flux freezing with the charged medium,
this transfer of Alfv\'enic power into the shear is dictated by the
diffusion rate within the resistive plasma in the region $|y|>s$,
where the resistance is provided primarily by $e^-$/radio photon
scatterings that occur with a frequency $\sim n_{ph} \,\sigma_{MC}\; c$,
in terms of the photon number density $n_{ph}$ and the magnetic Compton
cross section $\sigma_{MC}$.  With $\epsilon_0$ the characteristic
radio photon energy, each event imparts a momentum $\sim\epsilon_0/c$
to the electron (moving nonrelativistically with a velocity
$v_e\gg\epsilon_0/m_e c)$, which must therefore interact with
$\sim(m_e v_ec/\epsilon_0)(c/v_e)$ photons in order
to suffer a significant deviation to its path.
Thus, the electron deflection frequency is
$$
\nu_e \sim {F_r\,\sigma_{MC}\over m_ec^2} \approx 4.3\times 10^{11}\;\s^{-1}
\;\xi \;
\left({\eta\over 10^3}\right)^{-2}
\left({B_{a0}\over 10^{12}\;\G}\right)^2
\left({s\over 10 \;\cm}\right)^{-2}\left({P\over 5\;\s}\right)
,\eqno(5)
$$
where $\sigma_{MC}\approx \sigma_T/4$ when $B_{a0}\sim B_0\sim 10^{12}\;\G$
(Dermer 1990).

With a conductivity $\sigma\approx n_0\, e^2/m_e
\nu_e$ and a diffusion time scale
$\tau_d \equiv 4\pi\sigma S^2/c^2$
(the well known MHD value which is valid as long
as $\tau_d\ll 2\pi/\omega$), the diffusion
velocity $v_d\approx S/\tau_d$ for the
magnetic field lines is given as
$$\eqalignno{
v_d = &\hbox{min}\Bigl[c,8.7\times 10^8\; \hbox{cm s}^{-1}\cr
&\times\xi\;\left({\eta\over 10^3}\right)^{-2}\;
\left({B_{a0}\over 10^{12}\;\G}\right)^2\;\left({P\over 5 \,\s}\right)^2
\left({B_0\over 10^{12}\,\G}\right)^{-1}
\left({s\over 10\,\cm}\right)^{-2}\left({S\over 10^4\,\cm}\right)^{-1}
\Bigr]\;.&(6)}
$$
Thus, since $u_\alpha\approx c$, the waves dissipate over a lengthscale
$$\eqalignno{
R_d\equiv &\left({Sc\over v_d}\right)\approx
\hbox{max}\Bigl[S,3.5\times 10^5\,\cm\;\cr
&\times\xi^{-1}\;\left({\eta\over 10^3}\right)^{2}\;
\left({B_{a0}\over 10^{12}\,\G}\right)^{-2}\;
\left({P\over 5 \,\s}\right)^{-2}
\left({B_0\over 10^{12}\,\G}\right)\left({s\over 10\,\cm}\right)^2
\left({S\over 10^4\cm}\right)^{2}\Bigr]\;.&(7)}
$$
As long as $R_d\simless R_*$ (i.e., $\xi$ is
sufficiently large), most of the wave
energy is channeled into the shear before the
waves break, and we may equate the Alfv\'enic
luminosity ($\sim B_{a0}^2 c A_w/8\pi$)
generated at the stellar surface with the
power ($\sim E_{Az} J_s A_s R_d$) dissipated by
the current as it converts magnetospheric energy into
upscattered radiation.  Here,
$A_w$ and $A_s$ are the surface areas corresponding
to the wave and shear regions, respectively.
This yields an average electric field strength
$$\eqalignno{
E_{Az} \approx &\hbox{min}\Bigl[5\times 10^{11}\hbox{sV}\;\cm^{-1}\;
\left({B_{a0}\over 10^{12}\,\G}\right)
\left({\eta\over 10^3}\right)
\left({s\over 10\,\cm}\right)
\left({S\over 10^4\,\cm}\right)^{-1}
,1.4\times 10^{10}\,
\hbox{sV}\;\cm^{-1}\;\cr
&\times\xi\;\left({\eta\over 10^3}\right)^{-1}\;
\left({B_{a0}\over 10^{12}\,\G}\right)^{3}\;
\left({P\over 5\, \s}\right)^{2}
\left({B_0\over 10^{12}\,\G}\right)^{-1}
\left({s\over 10\,\cm}\right)^{-1}\left({S\over 10^4\,\cm}\right)^{-2}
\Bigr]\;.&(8)}
$$

As before, we assume a typical pulsar spectrum specified
as a steep power law with
(flux density) index $\mu$ above a break at
frequency $\epsilon_0/h\approx 500$ MHz.
With $\gamma\gg 1$, a lab frame photon with
energy $\epsilon$ will be blue-shifted
to $\sim 2\gamma\epsilon$ in the electron rest
frame, which is well below the resonant
energy $\epsilon_B\equiv (B_0/44.14\times 10^{12}\,\G)\;m_ec^2$,
and its angle of propagation
relative to the particle direction
(and hence $\b_0$) is $\sim \gamma^{-1}$.  As such,
$\epsilon_B/\epsilon \approx 6\times 10^9(B_0/ 10^{12}\,\G)
(\epsilon/h\,500\,\hbox{Mhz})^{-1} < \gamma^2$,
and $\sigma_{MC}\approx 4\sigma_T(\gamma\epsilon/\epsilon_B)^2$
(Melia \& Fatuzzo 1989; Dermer 1990).
Balancing the accelerating force $eE_{Az}$
by the radiative drag $\gamma^2 F_r \sigma_{MC}/c$,
one therefore obtains
$$
\gamma_s \approx 10^6
\left({B_{a0}\over 10^{12}\,\G}\right)^{1/4}
\left({B_0\over 10^{12}\,\G}\right)^{1/4}
\left({P\over 5 \,\s}\right)^{1/4}
\left({S\over 10^4\,\cm}\right)^{-1/4}
\left({\epsilon_0\over h\,500\,\hbox{MHz}}\right)^{-1/2}
\;f(\xi)\;,\eqno(9)
$$
where $f(\xi)=\hbox{min}[1,(\xi/\xi_0)^{-1/4}]$ and
the parameter $\xi_0$ is defined as the smallest value
of $\xi$ for which $v_d=c$ (e.g., $\xi_0=35$ for the represented
parameter space), and where we have used $\eta=S/s$.
This result is consistent with the assumptions discussed above
(e.g., that the particle motion is relativistic and sufficiently
energetic to produce the required radio luminosity and that $\sigma_{MC}$
have an $\epsilon^2$ dependence).
Ultimately, the current $J_s$ must decay in concert
with the Alfv\'enic magnetic field, even
though the particles remain relativistic.
Evidently, the initially fully charge separated regions must merge
together and neutralize.
This behavior is expected since the power
transferred from the wave to the particles is reduced as the
magnetic field decays. As such, an increasing number of charges
undergoing collisions will not be energized quickly enough to
remain in phase with the wave, and are therefore
swept up by the lagging oppositely charged wave region.

The scaling of the magnetic fields
$B_0$ and $B_{a0}$ in the above equations was chosen for convenience
and not to suggest that $B_{a0}\sim B_0$.
We note, however, that the presence of
significant reconnection in the nonlinear regime
would have the desirable effect of enhancing the annihilation rate
{\it within} the shear (see the paragraph preceeding Eq. 5).

\medskip
\centerline{\bf 3. The Cosmological Radio Pulsar Model For Gamma-ray Bursts}
\medskip
In applying the above discussion to the cosmological
gamma-ray burst model, we must now generalize
to a more realistic magnetospheric geometry in which the
field lines are more or less radial close
to the stellar surface (cf. Eq. 7).  Although globally the results
of \S 2 are expected to apply here, we note an important difference
between the two geometries.  The underlying magnetic field
strength $B_0$ now decreases away from the stellar surface as $(R_*/r)^2$,
whereas in the absence of wave damping, the Alfv\'enic field decreases
only as $(R_*/r)$, for which the waves eventually lose their
linearity and shock.  However, since the wave energy
diffuses into the shear on a length scale $<R_*$,
the Alfv\'enic field also decreases roughly as $(R_*/r)^2$.

It is evident from
\S 2 that SAWs focus their energy into the
internal current flow. Thus, as long as $\gamma^{-1}\ll s/R_*$,
the flux is enhanced by a factor
$\eta$ in certain directions, for which a source at
a distance $D$ will have an observable
$\gamma$-ray flux
$$
F_\gamma = \eta {B_{a0}^2\over 8\pi} c \left({R_*\over D}\right)^2
=  1.3 \times 10^{-7}\, \ergs\;\cm^{-2}\;\s^{-1}\left({\eta\over 10^3}\right)
\left({B_{a0}\over 10^{12}\;\G}\right)^2
\left({D\over 1\,\hbox{Gpc}}\right)^{-2}.
\eqno(10)
$$
The observability of these bursts at cosmological distances
imposes strict (but not unrealistic)
conditions on the model parameters, such as
the required burst power ($L_{burst}\sim 10^{44}-10^{46}$ ergs s$^{-1}$),
whose magnitude depends on whether the SAWs
are generated only near the polar cap
(where the field lines are most strongly coupled to
toroidal crustal activity),
or are generated throughout the entire stellar surface.

Since the particle flow remains optically thin to the
radio photons impinging upon the star,
the $\epsilon^2$ dependence of the cross-section (see above) implies
that the incipient radio spectrum is upscattered to
a $\gamma$-ray spectrum with (power) index
$\mu+2+1$, and very importantly, that the spectral radio break at $\epsilon_0$
is translated to the corresponding $\gamma$-ray break at
$$
\epsilon_{break}\sim 2\gamma_s^2\epsilon_0 \approx 4.4\;\MeV
\left({B_{a0}\over 10^{12}\,\G}\right)^{1/2}\left({B_0\over
10^{12}\,\G}\right)^{1/2}
\left({P\over 5\, \s}\right)^{1/2}\left({S\over 10^4\,\cm}\right)^{-1/2}
\;g(\xi)\;,\eqno(11)
$$
{\it independent of $\epsilon_0$} (we have again used $\eta=S/s$).  Here,
$g(\xi)=\hbox{min}[1,(\xi/\xi_0)^{-1/2}]$.
This result compares favorably with the observed value of
$\epsilon_{break}$ (which after redshift is
taken into account is seen to fall within the range
$\sim 100$ keV $- 3$ MeV; Schaefer et al. 1992), and
suggests that $\xi/\xi_0$ may be as large as 100.
A more detailed description of the resulting $\gamma$-ray
spectrum is given in MF (see, for example, Figure 2 therein).
For  completeness, we note that a cylindrical
shear would correspond to $\eta\sim (S/s)^2$
and the parameter $S$ in
Equation (11) should be replaced by $s$.
Such a strictly confined shear region would thus
appear to be unlikely, though it cannot
be ruled out without a more detailed calculation.

The question of which mechanisms contribute to the
generation of SAWs is best addressed by considering
the constraints imposed on the model by the different
probabilities of observing extragalactic
and galactic events.  If the source distribution
is cosmological, the BATSE data imply a detection
rate ${\cal R}_{c} \approx 6\times 10^{-7}$ bursts
per year per galaxy, with an actual rate
${\cal R}_T = {\cal R}_{c}/P_c$, where $P_c$
is the probability of detecting the extragalactic burst.
Assuming that the number of galactic sources is roughly
equal to the average number of sources in
all other galaxies, we should therefore expect
to see a galactic rate ${\cal R}_g=R_TP_g$, where
$P_g$ is the corresponding probability of detecting
a galactic burst in progress.  According to
Equation (10), a galactic burst would be quite
distinguishable, exhibiting fluxes of order
$1$ ergs cm$^{-2}\,\s^{-1}$, i.e., $\sim 10^7$
times larger than their extragalactic counterparts.
Since none of these have ever been detected,
we infer that ${\cal R}_{g}\simless 0.01$ per year,
for which $P_c/P_g \ge 6\times 10^{-5}$.

What this means in practice is best seen with
recourse to a specific scenario.  Let us assume
that the underlying sources are turned-off pulsars
(i.e., neutron stars older than $\sim
10^7$ years), which have nonetheless
retained their strong magnetic
fields and have achieved alignment
at a period $P \sim 5\s$, corresponding to an open field line polar cap
radius $R_{pc}\approx 6\times 10^3 \cm$. The stellar rotation
implies that a given $\gamma$-ray flux region sweeps in and
out of the observer's view, but as long as the duration of an individual
sweep is longer than the instrument resolution
time $\tau_i$, the inferred (average) flux
is correctly given by Eq. (10). If, however,
the sweep time is shorter than $\tau_i$, the observed (average)
flux for the most distant bursts drops below the instrument sensitivity.
This means that an extragalactic burst is observable
predominantly within a cone centered about
the rotation axis, with an opening angle corresponding to
a radius $R_{ob} \equiv (s/\pi)(P/\tau_i)$ (two
poles).  Thus, since the probability that a sheared
region occurs within the area enclosed by this
``observable'' cap is $\sim R_{ob}/S$ (where $R_{ob} < S$),
the probability of seeing a burst in progress at cosmological
distances must be
$$
P_c\sim 7.6\times 10^{-10}  \left({s\over 10\,\cm}\right)^3
\left({P\over 5\,\s}\right)^3\left({\tau_i\over 64 \,\hbox{m}\s}\right)^{-3}
\left({S\over 10^4\,\cm}\right)^{-1}\;.\eqno(12)
$$
The actual event rate would therefore be $R_T\sim 10^3$
per galaxy per year, which in turn
implies a stellar repetition time scale of $\sim 10^3-10^6$
years if the population is comprised of
$\sim 10^6-10^9$ objects.

The probability of detecting such a burst in progress
within the galaxy is higher
since for these events $R_{ob}\sim R_{pc}$
(see above).  As such, $P_g$ is simply the ratio of solid
angles corresponding to the polar cap region and the
entire star (i.e., $4\pi$).  For
a $P=5\,\s$ rotator, we therefore infer
that $P_g\approx 2\times 10^{-5}$, so that
$P_c/P_g\sim 4\times 10^{-5}$ for this population, which is consistent with the
constraint discussed above.  To put this result in another way,
we would anticipate that such
a population of GRB sources should produce an
observable galactic ``super'' burst roughly once
every $\simgreat 50$ years. In addition, we note that GRBs
originating from extragalactic sources are expected to
be accompanied by $\sim 0.01-1.0$ Jansky
radio bursts (see MF).  For galactic bursts,
the observable radio flux will be roughly 10 orders of
magnitude larger and therefore (as is the
case for the $\gamma$-ray signal) quite distinguishable.
Assuming that the radio emission is produced
at $\sim 10 R_*$ above the polar cap, the probability of seeing
the radio burst will be roughly 10 times greater than of seeing
the corresponding $\gamma$-ray burst, suggesting an observable
galactic ``super'' radio burst rate of roughly once every
$\simgreat 5$ years.  This result does
not conflict with current observations since only a small fraction
of the sky is monitered by radio telescopes at any
given time.
A possible link between GRB sources
and soft $\gamma$-ray
repeater events, which in this picture would be
interpreted as bursts viewed outside of the
open field line cone, has been discussed elsewhere (Melia \& Fatuzzo 1993).

We are grateful to the anonymous referee for suggesting several
improvements to the manuscript.
This research was supported by NSF grant PHY 88-57218, and the NASA
High-Energy Astrophysics Theory and Data Analysis Program at UA, and by
the Compton GRO Fellowship Program at UM.
\vskip 0.6in
\centerline{\bf References}
\medskip
\ref Bhat, P. N. et al. 1992, {\it Nature}, {\bf 359}, 217.
\ref Canfield, E., Howard, W. M. \& Liang, E. P.\ 1987, \apj, {\bf 323}, 565.
\ref Dermer, C. D.\ 1990, \apj, {\bf 360}, 197
\ref Fatuzzo, M. \& Melia, F. 1993, \apj, {\bf 407}, 680.
\ref Ho, C. \& Epstein, R. I.\ 1989, \apj, {\bf 343}, 277.
\ref Meegan, C. A., et al. 1992, Nature, {\bf 355}, 143
\ref Melia, F.\  1990a, \apj, {\bf 351}, 601.
\ref Melia, F.\  1990b, \apj, {\bf 357}, 161.
\ref Melia, F. \& Fatuzzo, M.\ 1989, \apj, {\bf 346}, 378.
\ref Melia, F. \& Fatuzzo, M.\ 1992, \apjl, {\bf 398}, L85 (MF)
\ref Melia, F. \& Fatuzzo, M.\ 1993, \apjl, {\bf 408}, L9.
\ref Narayan, R., Paczy\'nski, B. \& Piran, T. \ 1992, \apjl, {\bf 395}, L83.
\ref Pozdnyakov, L. A., Sobol, I. M. \& Sunyaev, R. A.\ 1977,
{\it Sov. Astron.}, {\bf 21}, 6.
\ref Schaefer, B. E., et al. 1992, \apjl, {\bf 393}, L51.
\ref Tamblyn, P. \& Melia, F. 1993, \apjl, submitted.

\vfill\eject\null
\hbox{\bf Authors' Addresses:}
\vskip 0.3 in
\settabs\+\noindent&{\bf Marco Fatuzzo:}\quad&\cr
\+&{\bf Marco Fatuzzo:}&Department of Physics,\cr
\+&\null&The University of Michigan, Ann Arbor, MI 48109\cr
\+&\null&[fatuzzo@pablo.physics.lsa.umich.edu]\cr
\+&{\bf Fulvio Melia:}&Department of Physics and Steward Observatory,\cr
\+&\null&University of Arizona, Tucson, AZ 85721\cr
\+&\null&[melia@nucleus.physics.arizona.edu]\cr
\end